**Dynamics of the sub-ambient gelation and shearing of solutions of P3HT incorporated with a non-fullerene acceptor o-IDTBR towards active layer formation in bulk heterojunction organic solar cells**

Li Quan[1], Dongrun Ju[1], Stephanie Lee[1] and Dilhan M. Kalyon[1,2,*]
[1] Chemical Engineering and Materials Science
[2] Biomedical Engineering
Stevens Institute of Technology
Hoboken, NJ

* for correspondence: dkalyon@stevens.edu

**Abstract**

Organic solar cells (OSCs) containing an active layer consisting of a nanostructured blend of a conjugated polymer like poly(3-hexylthiophene) (P3HT) and an electron acceptor molecule have the potential of competing against silicon-based photovoltaic panels. However, this potential is unfulfilled primarily due to interrelated production and stability issues. The generally employed spin coating process for fabricating organic solar cells cannot be scaled up and alternatives, especially relying on continuous polymer processing methods like extrusion and coating, cannot be readily applied due to the typically low shear viscosity and elasticity of polymer solutions making up the active layer. Recently, He *et al.*, have reported that the gelation of P3HT with [6,6]-phenyl-C61-butyric acid methyl ester ($PC_{60}BM$) under sub-ambient conditions can provide a new route to the processing of organic solar cells and that increases in power conversion efficiencies (PCEs) of the P3HT/$PC_{60}BM$ active layer are possible under certain shearing and thermal histories of the P3HT/$PC_{60}BM$ gels. Here oscillatory and steady torsional flows were used to investigate the gel formation dynamics of P3HT with a recently proposed non-fullerene o-IDTBR under sub-ambient conditions. The gel strengths defined on the basis of linear viscoelastic material functions as determined via small-amplitude oscillatory shear were observed to be functions of the P3HT and o-IDTBR concentrations, the solvent used and the shearing conditions. Overall, the gels which formed upon quenching to sub-zero temperatures were found to be stable during small-amplitude oscillatory shear (linear viscoelastic range) but broke down even at the relatively low shear rates associated with steady torsional flows, suggesting that the shearing conditions used during the processing of gels of P3HT with small molecule acceptor blends can alter the gel structure and possibly affect the resulting active layer performance.

**Introduction**

Organic solar cells in theory represent serious potential alternatives to silicon-based solar panels that are widely employed worldwide [1. D. Wöhrle, 1991]. Fig. 1 shows an example of an OSC used to convert sunlight to electricity. The typical design of an OSC is shown in Fig. 2, which involves an anode, a cathode and two layers consisting of an electron donor polymer and an electron acceptor small organic molecule [2. S. Günes, 2007] [3. N. S. Sariciftci, 1992].

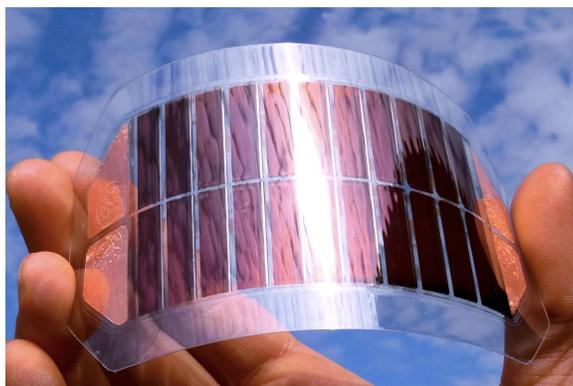

Figure 1. A flexible organic solar cell module [4].

In the simplest design, OSCs consist of two semiconducting layers on top of each other, i.e. a bilayer structure. During typical solar cell operation, photons excite electrons in the electron donor, leaving behind empty spaces in the highest occupied molecular orbital (HOMO) level, referred to as "holes". Excited electrons and holes are attracted to each other, constituting a tightly bound pair, i.e., the exciton. If the exciton can diffuse to the interface between the electron donor and electron acceptor layers before recombining, the exciton can dissociate by transferring the electron to the electron acceptor. Electrons (holes) then can travel through the electron acceptor (donor) to the cathode (anode) via charge hopping. Charge transfer is facilitated by a built-in electric field created by the different work functions of the anode and cathode [5. A. M. Bagher, 2015].

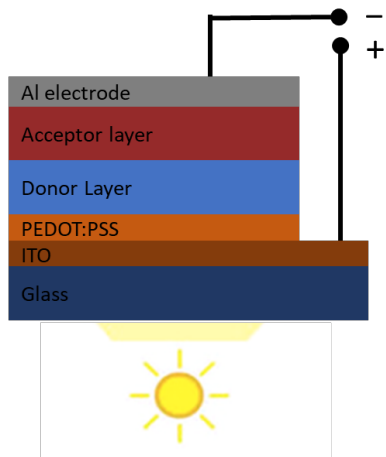

Figure 2. The layered design of an organic solar cell.

As Fig. 1 shows organic solar cells can be constructed to be thin, lightweight and flexible. As such, they are also easy to install. On the basis of these characteristics, organic solar cells have significant advantages over rigid, heavy and difficult to install silicon-based solar cells. Organic solar cells can be especially suitable in applications involving installations at the outsides of buildings and on the surfaces of complex shaped products, including vehicles, backpacks and tents [6. M. Jacoby, 2016]. Defense applications would have been a serious market for organic solar cells since they have the capability to be placed over the surfaces of military equipment. However, there are currently no commercially available organic solar cells and their potential is unfulfilled.

Another type of promising OSC is based on the bulk heterojunction design as shown in Fig. 3. In heterojunction cells the light-absorbing component is an active layer, i.e., a mixture, consisting of a conjugated polymer that is mixed intimately with an electronegative molecule, to constitute a nanostructured blend [7. A. Gusain, 2019]. The physics of basic heterojunction device configurations have been investigated extensively [7. A. Gusain, 2019] [8. C. J. Brabec, 2001] [9. E. Bundgaard, 2007] [10. D. Chen, 2011].

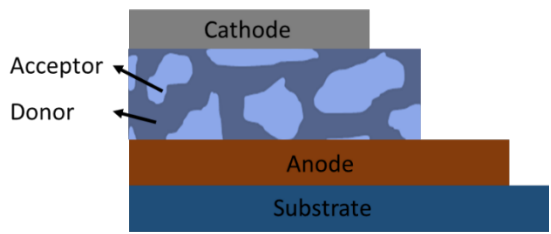

Figure 3. The modern bulk heterojunction cell design.

The most widely investigated donor polymer is the homo-polymer P3HT [11. S. Holliday, 2016] [12. C. J. Mulligan, 2014] [13. R. Po, 2014] and typical organic electron acceptor molecules are PC$_{60}$BM [14. N. S. Sariciftci, 1993], PC$_{71}$BM [15, M. M. Wienk, 2003], perylene diimides (PDI) [16. B. M. Savoie, 2014], FEHIDT [17. K. N. Winzenberg, 2013], etc. P3HT is a sulfur containing heterocyclic polymer, i.e., a polythiophene with a short alkyl group (hexyl, $CH_2(CH_2)_4CH_3$) on each repeat unit with a chemical formula of $(C_{10}H_{14}S)_n$. It is obtained via the polymerization of the monomer 3-hexylthiophene. The chemical structure is shown below:

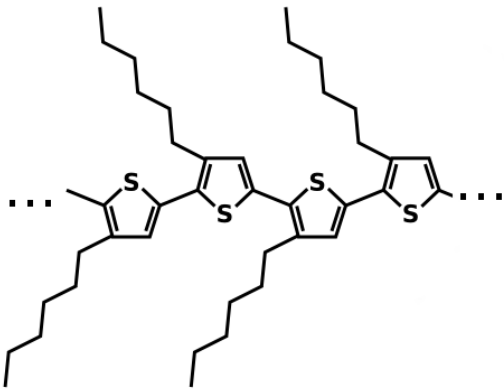

Figure 4. The chemical structure of poly(3-hexylthiophene) (P3HT).

P3HT can be found in two forms, regioregular P3HT (RR-P3HT) and regiorandom P3HT (RRa-P3HT), as shown below [18. M. Giulianini and N. Motta, Chapter 1, 2012], the difference is in the alternating positions of the alkyl chains along the backbone.

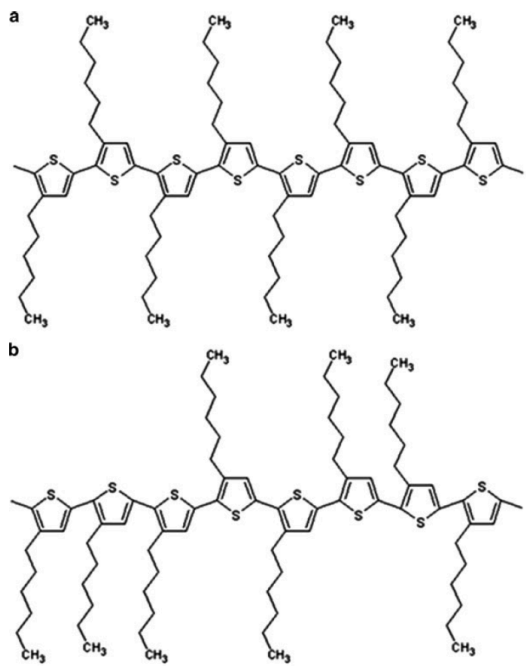

Figure 5. P3HT in the (a) regioregular form and (b) regiorandom form.

P3HT is relatively stable [19. M. Jørgensen, 2012] and can be synthesized readily [20. K. Tremel, 2014]. Several different methods are reported for synthesis of highly regioregular P3HT [21. M. Ansari, 2018] [22. M. Naito, 2008] [23. W. Yi, 2008] [24. J. H. Bannock, 2013]. P3HT with regioregular structure can aggregate and crystallize via $\pi$ (also called $\pi$-$\pi$ stacking) interactions [25. M. Bernardi, 2010] [26. M. Aryal, 2009] [27. J. Xiao, 2015] [28. V. Skrypnychuk, 2015]. $\pi$-stacking occurs due to attractive, noncovalent interactions between aromatic rings, since they contain $\pi$ bonds. The unit cell of RR-P3HT exhibits the unit cell dimensions of a=16.8 Å, b=3.8 Å and c=7.7 Å [29. M. Giulianini, 2012]. It has a density of 1,333 kg/m$^3$ [30. Y. SUN, 2013]. The crystal structure of RR-P3HT is shown below [18. M. Giulianini and N. Motta, Chapter 1, 2012]:

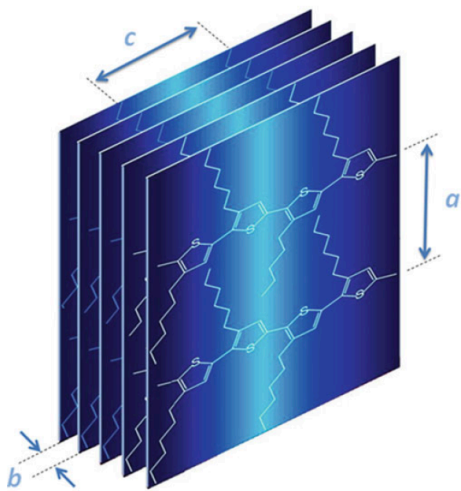

Figure 6. RR-P3HT in its crystalline form.

During the solution based fabrication of proto-type solar cells, RR-P3HT is dissolved in various solvents, including tetrahydrothiophene, chlorobenzene, 1,2-dichlorobenzene (o-DCB), o-xylene, with solubilities of 37.6, 15.9, 14.7, 2.7 mg/ml, respectively [31. N. Espinosa, 2014]. When dissolved, RR-P3HT is known to form colloidal particles in solution via crystallization. The formation of RR-P3HT colloidal particles and possibly crystalline nanofibers (also referred to as nanowhiskers) and their percolation, i.e., network formation via interactions with each other lead to the formation of volume-spanning networks. Such networks give rise to gelation, i.e., the formation of microgels, with junction points comprised of interacting crystal domains. Besides use in organic solar cells, RR-P3HT nanofibers have also been used for the fabrication of organic phototransistors (OPTs) [32. L. Persano, 2015].

Linear viscoelastic material functions characterized via small-amplitude oscillatory shearing are very sensitive to the micro and nanostructures formed upon crystallization of the polymer phase in polymer solutions. The rate of gelation and the gel strength can be characterized via small amplitude oscillatory shear experiments.

A number of investigations have focused on the fabrication of active layers for bulk heterojunction applications by blending P3HT with PC$_{60}$BM. In early investigations, the efficiencies of the P3HT/PC$_{60}$BM solar cell systems were around 2.5% [33. A. J. Moulé, 2014] [34. S. E. Shaheen, 2001]. The power efficiencies could be increased to 3.5% by 2003 [35. F. Padinger, 2003] and to 5% by 2005 [36. G. Li, 2005] [37. W. Ma, 2005]. More recently, the PCEs of devices based on P3HT/PC$_{60}$BM blends could be increased up to 11% upon doping with low concentrations of iron (II, III) oxide nano-particles (Fe$_3$O$_4$) [38. D. M. González, 2015].

One of the major objectives of organic solar device research has been to replace the fullerene acceptors used in heterojunction devices including PC$_{60}$BM with non-fullerene based electron acceptors. For example, PCEs of 15% and over 17% have been obtained using Y6 [39. J. Yuan, 2019] and ITIC [40. Y. Lin, 2015] acceptors, respectively. It was also shown that the optimization of device structure and film morphologies is essential to achieve such high PCEs. New non-fullerene acceptors like IDTBR (rhodamine-benzothiadiazole-coupled indacenodithiophene), for example, ((5Z,5′Z)-5,5′-(((4,4,9,9-tetraoctyl-4,9-dihydro-s-indaceno[1,2-b:5,6-b′]-dithiophene2,7 diyl)bis(benzo[c][1,2,5]thiadiazole-7,4-diyl))bis(methanylylidene))bis(3-ethyl-2-thioxothiazolidin-4-one)) (o-IDTBR) have also been proposed [11. S. Holliday, 2016] [41. N. Gasparini, 2017] [42. D. Baran, 2017] [43. S. F. Hoefler, 2018] [44. C. H. Tan, 2018] [45. Q. Liang, 2019] [46. J. I. Khan, 2019] [47. E. Pascual-San-José, 2019] [48. J. Liu, 2020] [49. K. An, 2020]. The chemical structure of PC$_{60}$BM and o-IDTBR is as follows:

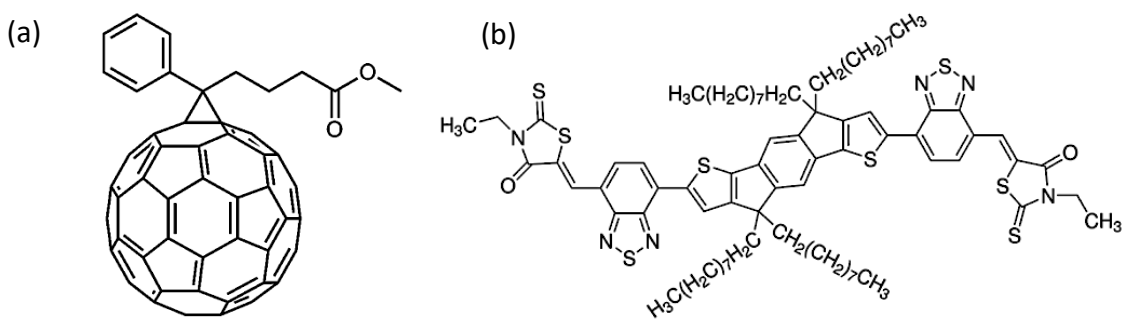

Figure 7. The chemical structure of (a) PC$_{60}$BM. (b) o-IDTBR.

The PCE window of the P3HT/o-IDTBR system was found to be in the range of 6.4 % to 7.8% [11. S. Holliday, 2016] [42. D. Baran, 2017] [45. Q. Liang, 2019] [46. J. I. Khan, 2019]. Holliday *et al.* could reach a PCE of 6.4% by annealing the system at 130°C. The use of a cosolvent 1,2,4-triclorobenzene (TCB) provided incremental increases in fill factor to 0.70 and PCE to 7.18% [45. Q. Liang, 2019]. Baran *et al.* showed that introducing a third component, namely o-IDFBR, into the P3HT/O-IDTBR blend increased the PCE close to 8% [42. D. Baran, 2017]. Overall, what made the P3HT/o-IDTBR system very interesting was the observed enhancement of the stability of the solar devices manufactured with this donor/acceptor combination.

The typical solution processing method used in fabricating the active layer of the bulk heterojunction solar cells is spin coating. Spin coating involves the placement of the electron donor/acceptor blend onto a disc which is typically rotated up to 5000 rpm to provide typically a 100-150 nm thick coating on indium tin oxide (ITO) glass substrates to constitute the active layer. However, spin coating is a batch process and is thus difficult to scale up. Alternatives to spin coating can be considered on the basis of various polymer mixing and film-forming processing techniques such as extrusion followed by a coating method, including doctor blading, casting, slot-die coating, gravure coating, knife-over-edge coating, off-set coating, spray coating and printing techniques such as ink jet printing, pad printing and screen printing [50. F. C. Krebs, 2009].

However, the shear viscosity and elasticity of the polymer electron donor/organic acceptor blend solutions under typical ambient temperature conditions should be appropriate for the application of such well-known polymer processing methods like extrusion and coating. For example, if the rheological behavior of the blend were to be suitable, twin screw extrusion and coating could have been used to intimately blend the polymer phase with the electron acceptor and solvents. Such intimate mixing processes were demonstrated for many blends of complex fluids whereby the goodness of the mixing of the ingredients, mixing indices [51. D. M. Kalyon, 2006] [52. M. Erol, 2005] [53. R. Yazici,1993], could be controlled through the manipulation of the thermo-mechanical history imposed during extrusion [51. D. M. Kalyon, 2006] [52. M. Erol, 2005]. The thermo-mechanical history during processing can be mathematically modeled and the simulation results can be used to optimize twin screw extrusion geometries and conditions [54. A. Lawal, 1995] [55. D. M. Kalyon, 2007] [56. M. Malik, 2005] [57. M. Malik, 2014] and roll based processes like continuous shear roll milling [58. D. M. Kalyon, 2004].

Thus, if the low viscosity and elasticity problem of the blends can be overcome there is great potential for the use of relatively cheap and scalable extrusion/coating based methods for the fabrication of the active layers of bulk heterojunction solar cells. A combination of twin screw extrusion and coating apparatus that is assembled in our laboratories is shown in Fig. 8.

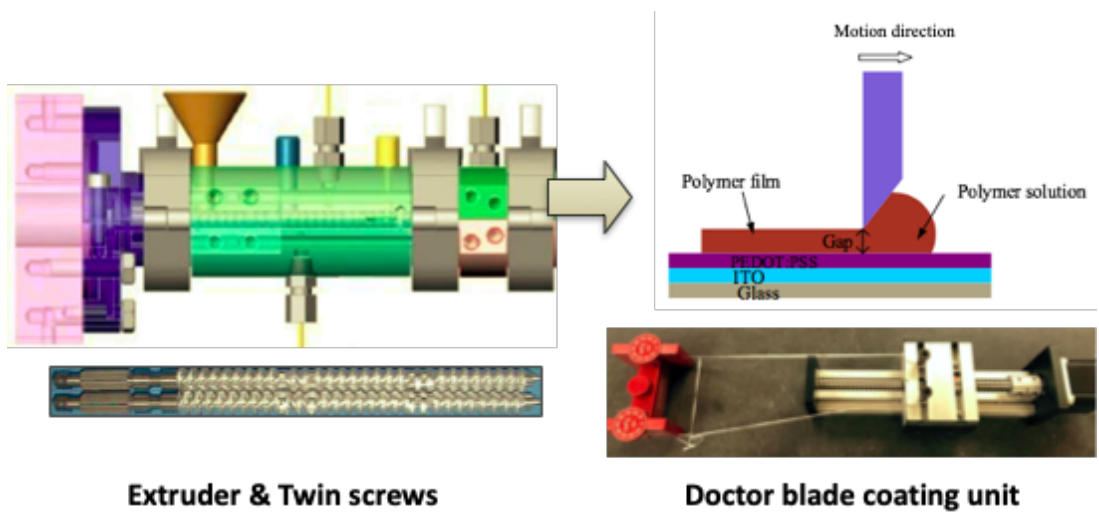

Figure 8. Twin screw extrusion slide with coating unit.

Overall, it is well known that the nanostructures that are generated as a result of the thermo-mechanical history that is applied during the process significantly affect the efficiencies of the devices that are fabricated [7. A. Gusain, 2019]. Thus, the challenges are two-fold. First, the rheological behavior and processability of the donor/acceptor blend system need to be tailored to be suitable for continuous processing, and second the processing conditions and geometries need to be optimized during the extrusion/coating methods to generate nanostructures with improved stabilities and power efficiencies.

Recently, He *et al.*, have demonstrated the formation of porous crystalline networks via viscoelastic phase separation (VPS) [59. H. Tanaka, 2000] [60. H. Tanaka, 2006] [61. H. Tanaka, 2009] of rapidly cooled P3HT solutions. The key to viscoelastic phase separation is the asymmetry in molecular dynamics between the two components of the solution [59. H. Tanaka, 2000]. In a polymer solution, the polymer phase relaxes much more slowly compared to the surrounding solvent molecules. More explicitly, the slow relaxing fluid component (polymers) cannot catch up with the fast relaxation of the viscous solvent during deformation. This "viscoelastic phase separation" (VPS) followed by crystallization of P3HT has suggested a new route for a promising and scalable method to achieve optimized morphologies in the active layers of P3HT/PC$_{60}$BM based organic solar cells [62. J. He, 2019]. He *et al.* indicated that: "… upon rapid cooling, viscoelastic phase separation occurring in electron-donating RR-P3HT solutions leads to the formation of stable, crystalline RR-P3HT gel networks with hierarchical porosity that improve the light harvesting efficiency of solar cell active layers" [62. J. He, 2019]. This crystallization-induced gelation during VPS was found to be thermo-reversible and insensitive to the presence of both PC$_{60}$BM and noncrystallizing Rra-P3HT [62. J. He, 2019]. It was proposed that once formed in solution, these semicrystalline RR-P3HT networks can be transferred to OSC device platforms via continuous processing including, doctor blading. It was revealed that OSCs comprising cooled photoactive layers displayed 45% higher efficiencies compared to those comprising uncooled photoactive layers.

The objective of the current investigation is to build upon the earlier investigation of He *et al.* by characterizing the sub-ambient gel formation behavior of P3HT blended with o-IDTBR, i.e., another popular small molecule acceptor in two solvents and as a function of concentrations of P3HT and o-IDTBR, using rheological analysis as a tool and to compare the rheological behavior and resulting gel formation dynamics with those of P3HT blends with PC$_{60}$BM. Thus, this follow up investigation was carried out on a non-fullerene acceptor in comparison to the fullerene based small molecule acceptor of the He *et al.* investigation. As noted, Holliday *et al.* have shown that the o-IDTBR acceptor has significant advantages over the conventional PC$_{60}$BM acceptors due to the improved shelf-life and photo-oxidation stability of the P3HT/o-IDTBR system in comparison to the conventional P3HT/PC$_{60}$BM system [11. S. Holliday, 2016].

**Experimental Section**

**Materials.** In this investigation we have used only regioregular P3HT, i.e., RR-P3HT. Thus, all subsequent references to P3HT should be taken to mean RR-P3HT. P3HT was purchased from Rieke Metals (Lincoln, NE), with weight-average molecular weight of 50−70 kDa and regioregularity=91% (the degree of regioregularity was measured through the number of head-to-tail linkages in the polymer). o-IDTBR was procured from Derthon, Inc. (Shenzhen, CN) with purity>99% and Mw=1326.03 g/mol. The solvents o-DCB and o-xylene (purity≥99%) were received from Sigma-Aldrich. All materials were used as received without further purification.

**Sample Preparation.** Solutions of pure P3HT and blends of P3HT/o-IDTBR were prepared by dissolving P3HT in o-DCB or o-xylene on top of a hot plate under magnetic stirring conditions. The rotational speed of the magnet was set at 500 rpm and the temperature of dissolution was 70 °C. Typically, the dissolution took about 2-3 hours. Total concentrations of P3HT in the solvent were varied between 6.25 to 25 mg of P3HT per ml of solvent, i.e., mg/ml. Solutions of blends of P3HT and o-IDTBR were prepared at the P3HT over o-IDTBR weight ratios of 1:1, 1:1.6 and 1:2.

The rheological characterization of the P3HT and blends of P3HT/o-IDTBR solutions was carried out at the sub-ambient temperature of −5±0.2 °C using small-amplitude oscillatory shear and steady torsional shear flows. Two sets of solutions in o-DCB and o-xylene were prepared and characterized. The temperature used, i.e., −5±0.2 °C is the same temperature that was used earlier by He *et al.* in their experiments on P3HT/PC$_{60}$BM blends, the shearing of which under sub-ambient conditions provided benefits for increasing the conversion efficiency of photo-voltaic devices fabricated from P3HT/PC$_{60}$BM [62. J. He, 2019].

An Advanced Rheometric Expansion System (ARES) rotational rheometer available from TA Instruments of New Castle, DE was used in conjunction with a force rebalance transducer, 0.2K-FRTN1, and stainless steel parallel and cone and plate fixtures, all at a diameter of 8 mm. The cone angle was 0.1 rad. The torque accuracy of the transducer was ±0.02 g$_f$-cm. The actuator of the ARES is a DC servo motor with a shaft supported by an air bearing with an angular displacement range of $5×10^{-6}$ to 0.5 rad, and an angular frequency range of $1×10^{-5}$ to 100 rad/s. The ARES rheometer has an angular velocity range of $1×10^{-6}$ to 200 rad/s. The rheometer is equipped with an environmental control chamber, which can operate between −70 to 600 °C with a cooling and heating ramp rate from 0.1 to 50 °C/min.

**Small-amplitude oscillatory flow.**     During oscillatory shearing, the shear strain, $\gamma$, varies sinusoidally with time, *t*, at a frequency of $\omega$, i.e., $\gamma(t) = \gamma^0 sin(\omega t)$, where $\gamma^0$ is the strain amplitude. The shear stress, $\tau(t)$, response of the fluid to the imposed deformation consists of two contributions associated with the energy stored as elastic energy and energy dissipated as heat, i.e., $\tau(t) = G'(\omega)\gamma^0 sin(\omega t) + G''(\omega)\gamma^0 cos(\omega t)$, where the first part on the right of the equation is associated with the elastic energy storage and the second part is associated with dissipation. The two moduli, i.e., the storage modulus, $G'$, and the loss modulus $G''$, were characterized as a function of frequency, $\omega$. In the linear viscoelastic region, all dynamic properties are independent of the strain amplitude, $\gamma^0$, thus strain amplitude scans were applied first to assure that the dynamic properties could be determined under linear viscoelastic conditions. The use of a range of frequencies during oscillatory shearing provides the ability to fingerprint the characteristic time dependence of the linear viscoelastic properties. Considering that the Deborah number is the ratio of the relaxation time of the fluid over the characteristic time for the deformation, such frequency dependent experiments provide data over a broad range of Deborah numbers. Overall, two types of sinusoidal experiments were carried out in oscillatory shearing, i.e., time sweeps during which the frequency and the strain amplitude were kept constant upon reaching −5 °C and frequency sweeps that were imposed when asymptotic behavior was observed from the time sweeps carried out at −5 °C. The following provides the procedures that were followed.

During the experiments at −5 °C, the zero gap (plates touch each other at the temperature of the experiment) was set first. The desired gap separation between the two discs was then imposed at the same temperature. Upon setting the gap, the temperature was decreased to the targeted sub-ambient temperatures, −5 °C. Solutions were stirred on a hot plate at 70 °C, a predetermined volume was taken as a specimen (slightly greater than necessary to fill the gap) and via a pipette was transferred into the gap between the two discs and upon reaching the desired temperature the excess material was trimmed from the edge of the specimen. This process took about 1 min after the loading the specimen into the gap between the two discs. For each test, a fresh sample was loaded onto the rheometer to avoid any thermal and pre-shearing effects, unless indicated differently.

In order to characterize the sol-gel transition kinetics, oscillatory shear data were collected first as a function of time at constant $\gamma^0$ and $\omega$. Time sweep tests were stopped upon reaching of ultimate plateau values for $G'$ and $G''$ that could be sustained for durations longer than 10 min. The achievement of the plateau region, in dynamic properties indicates that the gel that was formed has achieved a steady-state structure, and that the solvent loss is negligible. Upon reaching the plateau region, the linear viscoelastic dynamic properties, $G'$ and $G''$, were collected as a function of frequency, $\omega$ at a constant $\gamma^0$. As noted earlier at the linear viscoelastic region the moduli are independent of the strain amplitude, $\gamma^0$.

**Steady torsional flow.** Steady torsional experiments were carried out in an attempt to characterize the development of the shear viscosity of the solution samples upon the imposed thermal treatment (exposure to $-5$ °C). These experiments were carried out using parallel plate fixtures employing multiple gaps. Multiple gaps are recommended to be used for complex fluids, including microgels and concentrated suspensions to be able to characterize the wall slip versus the shear stress relationship and to enable the determination of the true shear rates that are imposed [63. U. Yilmazer, 1989] [64. D. Kalyon, 1993] [65. B. K. Aral, 1994] [66. D. M. Kalyon, 2003] [67. D. M. Kalyon, 2005] [68. S. Aral, 2014] [69. J. F Ortega-Avila, 2016] [70. E. F. Medina-Bañuelos, 2017].

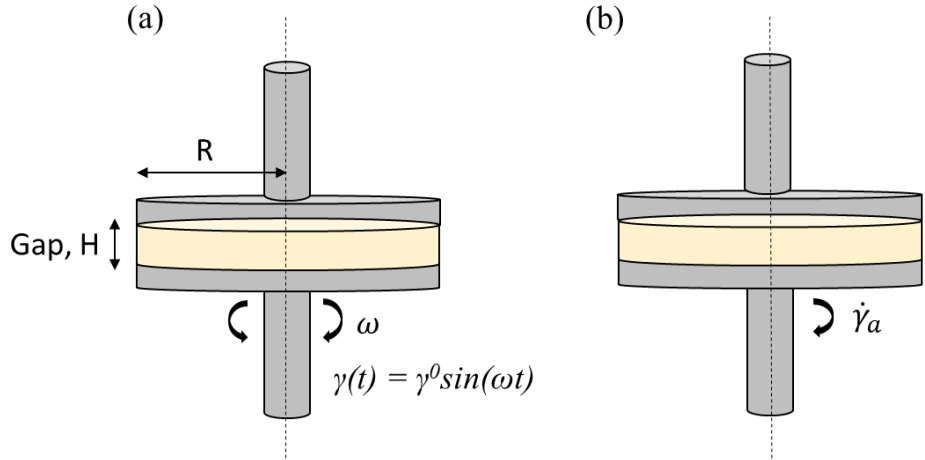

Figure 9. (a) Small-amplitude oscillatory shear. (b) Steady torsional shear.

The apparent shear rate, $\dot{\gamma}_a$ (not corrected for slip effects), is given by, $\dot{\gamma}_a = \left(\Omega r / H\right)$, where $r$ is the radial distance from the center of the disk, $H$ is the gap height, and $\Omega$ is the angular velocity of the bottom plate. The shear stress at the edge of the disk, $\tau_w(R)$, can be determined from $\tau_w(R) = \frac{\Im}{2\pi R^3}\left(3 + \frac{d\ln\Im}{d\ln\dot{\gamma}_{\alpha R}}\right)$, where $\Im$ is the torque generated by rotating the bottom disk, $\dot{\gamma}_{\alpha R}$ is the apparent shear rate at the edge of the disk. If plots of apparent shear rate versus reciprocal gap are drawn at constant shear stress at the edge, the slopes would be equal to $2U_s$, where $U_s$ is the slip velocity. The steady torsional flow experiments were carried out by systematically varying the surface to volume ratios via changes in the gap between the two parallel plates, i.e. 0.4, 0.8, and 1.2 mm.

**Differential scanning calorimetry (DSC) experiments**.   DSC experiments were carried out using a TA Instrument DSCQ100 at a heating rate of 5 °C/min under nitrogen. Samples were prepared by placing solutions into the ARES chamber at −5 °C for 5 min and then drop casting the solutions onto glass substrates. The solvent was then allowed to evaporate in a nitrogen glove box for 48 hr. After the solvent completely evaporated, the dried film was transferred  into DSC pans, constructed out of Al. No pre-shearing was applied, thus the rheometer was only used as a thermal treatment apparatus prior to the removal of the solvent under ambient temperature conditions (23±2 °C). The results of both microscopy and DSC experiments are included as supplemental files.

## Results and Discussion

The first set of figures, i.e., Fig. 10-13 show the dynamic properties of pure P3HT solutions at −5 °C. In these experiments P3HT solutions were cooled from room temperature to −5 °C and kept at −5 °C while subjecting the solution samples to small-amplitude oscillatory shear and recording the dynamic properties as a function of time. These time scans were then followed by frequency scans as explained below.

Fig. 10 shows the time dependencies of the $G'$ values of P3HT solutions in o-DCB at two P3HT concentrations of 12.5 and 25 mg/ml. These samples were quenched to −5 °C and the oscillatory shear was imposed at a frequency of 1 rad/s and a strain amplitude of 1% at the edge. At the P3HT concentration of 12.5 mg/ml the storage modulus grew slowly to reach a plateau value, i.e., about 800 Pa after about 3600 s. The 95% confidence intervals determined according to Student's t-distribution are presented and the confidence intervals are rather broad. The increase of the P3HT concentration to 25 mg/ml gave rise to an accelerated growth of the storage modulus which reached to a plateau of around 10 kPa within about 1500 s. The slope $dG'/dt$ (shown in the inset of Fig. 10) increases up to a maximum at 620 s and then decreases to zero after 1500 s at which the plateau region is reached.

At the plateau region the elastic moduli no longer changed with additional time of oscillatory shearing. The time to reach steady state, i.e., the reaching of the plateau region decreased as the concentration was increased. The time to reach the plateau decreased with increasing concentration of P3HT, i.e., from a duration of 3600 s at 12.5 mg/ml to 1500 s at 25 mg/ml. The $G'$ value at the plateau, denoted as $G'_{max}$, depended on the concentration of the P3HT, with the value of $G'_{max}$ increasing from 800 Pa to $10^4$ Pa as the P3HT concentration in o-DCB was increased from 12.5 to 25 mg/ml. The two picture insets of Fig. 10 show that the solutions were free flowing at room temperature and became solid-like upon reaching the plateau behavior.

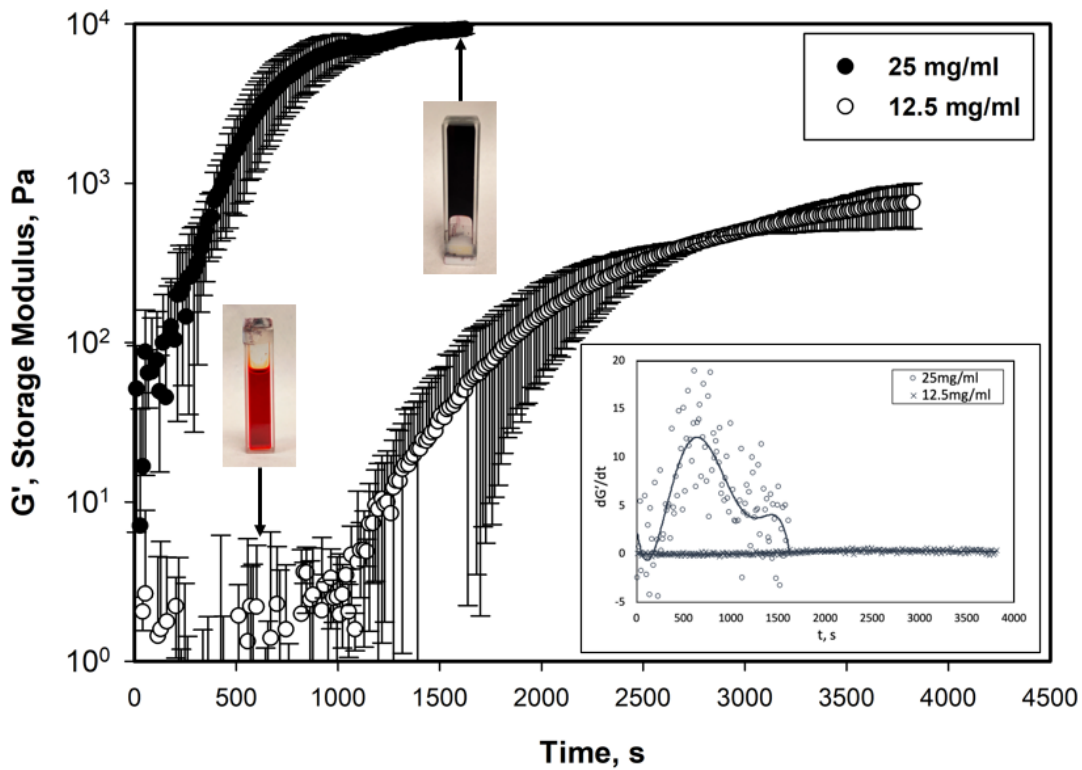

Figure 10. Storage modulus, $G'$, versus time, $t$, of P3HT in o-DCB with P3HT concentration of 12.5 and 25 mg/ml at −5 °C.

Fig.11 shows the dependency of the storage moduli, $G'$, on the frequency of oscillation in the plateau region for six concentrations of P3HT. These frequency sweeps were carried out following the reaching of the plateau behavior during time sweeps at 1 rad/s for 1000 s. The typical 95% confidence intervals determined according to Student's t-distribution are reported in Fig. 11. The confidence intervals obtained from three successive runs are relatively narrow indicating that the structures achieved during different runs at −5 °C were very reproducible to give rise to similar elasticity, as reflected in the narrow ranges of $G'$ obtained. It appears that the dynamic properties characterized in the linear viscoelastic range are stable to changes in frequency and time ranges associated with the relatively modest strains and strain rates of the linear viscoelastic range, indicating that the gels which form conserve their structures during subsequent oscillatory shearing in the linear range.

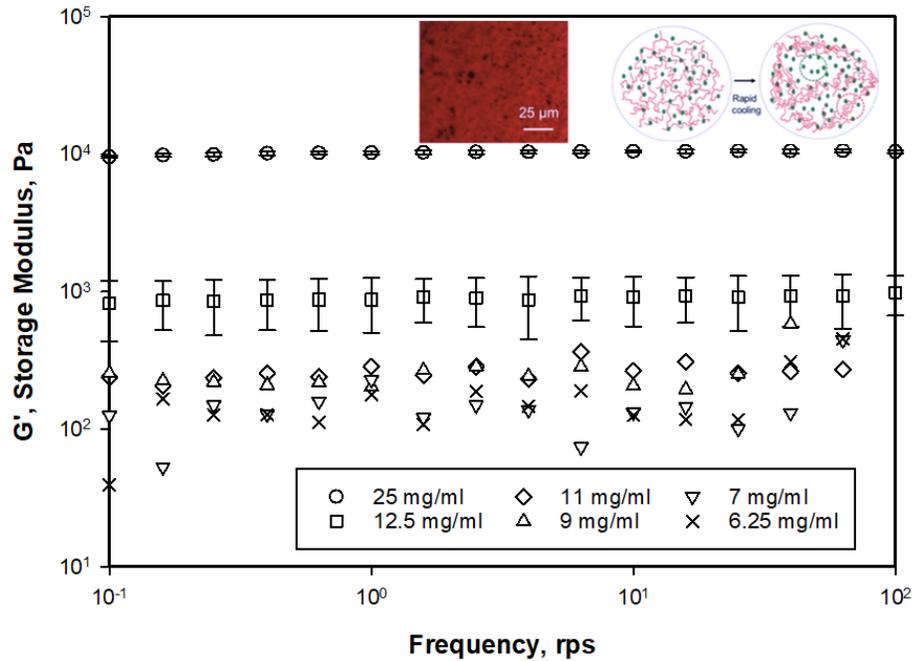

Figure 11. Storage modulus, $G'$, versus frequency, $\omega$, of P3HT solutions in o-DCB with concentrations of 6.25, 7, 9, 11, 12.5, and 25 mg/ml at −5 °C.

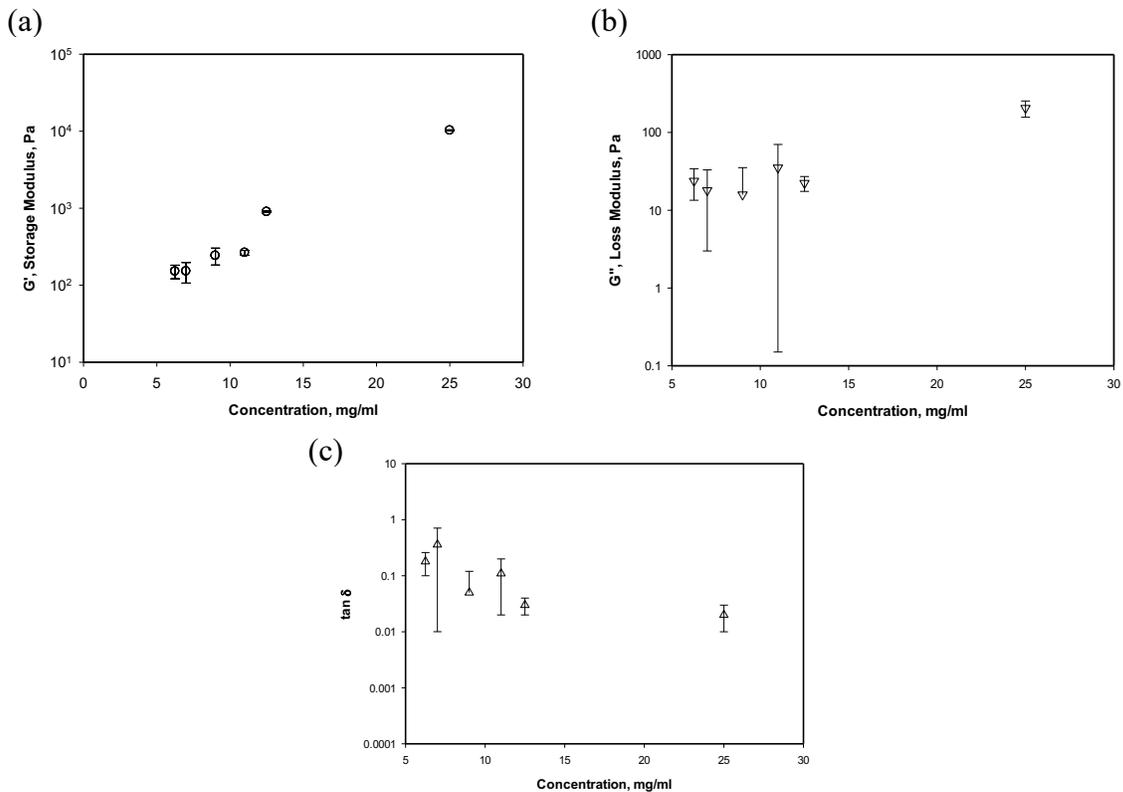

Figure 12. (a) Storage modulus, $G'$ (b) Loss modulus, $G''$ (c) tan $\delta$ versus different concentrations of 6.25, 7, 9, 11, 12.5, and 25 mg/ml of P3HT solutions in o-DCB with concentrations at −5 °C..

In the six sets of experiments reported in Fig.11 the concentration of the P3HT was varied between 6.25 to 25 mg/ml. At all concentrations, the storage moduli were independent of the frequency in the plateau region. Such independence of the dynamic properties from the frequency of oscillation in oscillatory shearing is one of the traits of "gel-like" behavior [71. H. H. Winter, 1986]. With increasing concentration of the P3HT the storage modulus values increase from 170 Pa at 6.25 mg/ml to about 800 Pa at 12.5 mg/ml. The frequency-independent storage modulus values represent the "gel strength" of the quenched P3HT in o-DCB samples. As shown in Fig. 11 the concentration of P3HT makes little difference on the gel strength in the concentration range of 6.25 to 11 mg/ml, with the storage moduli falling in the 170-270 Pa range. However, the concentration of the P3HT makes a significant difference in the storage modulus values, representing the gel strengths, when the concentration of P3HT was increased further. Going from 12.5 mg/ml to 25 mg/ml the storage modulus values increased from 800 to $10^4$ Pa. This behavior is consistent with the results of the earlier investigation of He *et al.* Their results indicated that when the concentration of P3HT in o-DCB was increased from 12.5 to 50 mg/ml the storage modulus of the gel increased from 1 to 50 kPa [62. J. He, 2019].

The inset in Fig. 11 shows a typical confocal fluorescence micrograph of cooled solutions of P3HT [62. J. He, 2019]. The cooling-induced structural rearrangements were associated with a viscoelastic phase separation process followed by crystallization in conjunction with the formation of a network of nanowhisker shaped crystallites [72. J. Liu, 2009]. The confocal micrographs obtained by He and co-workers of P3HT cooled to −5 °C revealed micron-sized solvent-rich "holes" (indicated as the darker regions in the micrograph shown in Fig. 11) due to phase separation and interchain crystallization as depicted schematically in the inset of Fig. 11. Cryogen based scanning electron micrographs revealed further an inter fibrillar network exhibiting nano sized pores [62. J. He, 2019]. Similar processes associated with interchain crystallization are also expected for P3HT and o-IDTBR blends resulting in the gelation behavior reported here.

Fig. 12a and b indicate that the $G'$ values are more sensitive to the structure of the gel in comparison to $G''$, i.e., $G'$ is dependent on the concentration of the P3HT for both the relatively small concentrations (6.25 to 12.5 mg/ml) as well as the higher concentration range of 12.5 to 25 mg/ml (Fig. 12a). Fig. 12b indicates that $G''$ is sensitive to the concentration of the P3HT only in the concentration range of 12.5 to 25 mg/ml. The results in Fig. 12 show that the elasticity of the gel (as represented by the values of $G'$) is more sensitive to concentration in comparison to the dependence of the viscosity and the viscous dissipation behavior of the gel versus the concentration of P3HT (as represented by the values of $G''$). Finally, tan $\delta = G''/G'$ values exhibit significantly broad confidence intervals and do not provide any clear cut dependence on P3HT concentration. The fact that both $G'$ and $G''$ increase with concentration may be a contributor, so that the $G''/G'$ ratios do not vary significantly with P3HT concentration.

Fig. 13 displays both the storage, $G'$, and the loss moduli, $G''$, as well as the tan $\delta$ values of pure P3HT solutions as a function of frequency in the plateau regions for two different solvents, o-DCB and o-xylene. . The results with both solvents again point to the gel-like behavior of the solution samples of P3HT upon reaching plateau regions following time scans of small-amplitude oscillatory shearing at −5 °C. First, the loss moduli, $G''$, are within a narrow range of 100 to 450 Pa and are independent of the frequency for both o-DCB and o-xylene. Consistent with the earlier observation that in o-DCB the storage moduli of the P3HT solutions are independent of frequency, the solution samples in o-xylene are also independent of frequency, i.e., they are within a small range of 9600 to 10500 Pa for o-xylene. For both solvents the storage moduli are significantly higher than the loss moduli ($G' >> G''$) over the entire frequency range tested, i.e., 0.1 to 100 rad/s. The $G'$ values are about 24-96 times greater than $G''$ values at similar frequencies. As shown in Fig. 12, the resulting values of tan $\delta = G''/G'$ are all smaller than 0.1, providing another manifestation of gel-like behavior [71. H. H. Winter, 1986]. Thus, for both solvents quenching the P3HT solution samples to $-5$ °C result in the formation of a gel. The gel strength values depend on the solvent used. Solution samples with o-DCB gave rise to significantly higher gel strengths (an order of magnitude higher $G'$ values for o-DCB versus o-xylene) in comparison to the gel strength of P3HT solutions with o-xylene. The loss modulus values were not affected by the solvent type.

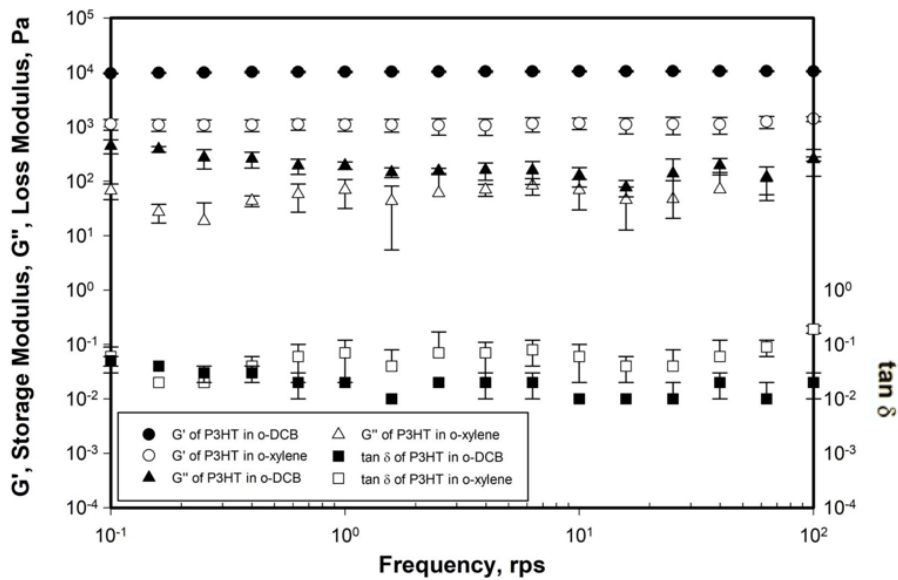

Figure 13. Storage modulus, $G'$, loss modulus, $G''$, and tan $\delta$ versus frequency, $\omega$, for 25 mg/ml P3HT in two different solvents o-DCB and o-xylene.

The next set of three figures, i.e., Fig. 14-16 show the dynamic properties of blends of P3HT with o-IDTBR. The time dependent behavior of the storage modulus for solution samples with equal amounts of P3HT and o-IDTBR (1:1 P3HT:o-IDTBR, 25 mg/ml each) dissolved in o-DCB and o-xylene when quenched to and kept at $-5$ °C at a constant frequency of 1 rad/s and a constant strain amplitude of 1% are shown in Fig. 14. Consistent with the earlier observed time-dependent behavior of pure P3HT solutions at $-5$ °C, the storage moduli of blends of P3HT with o-IDTBR also monotonically increase to asymptotic ultimate values, i.e., reach plateau values. With increasing oscillatory shearing time at $-5$ °C the storage modulus, which is indicative of the elasticity of the solution increased by about three orders of magnitude for both solvents at the equal P3HT and o-IDTBR concentration of 25 mg/ml.

Consistent with the behavior of pure P3HT (as was shown in Fig. 10) the insets included in Fig. 14 indicate that the P3HT and o-IDTBR blend solutions were free flowing at room temperature and became solid-like upon reaching $-5$ °C (when the container was flipped there was no flow).

The two solvents generated different time-dependent behavior. With o-DCB the storage modulus, $G'$ is relatively low at the beginning of the time sweep, followed by a significant increase with time of oscillatory shearing to reach an asymptotic plateau, i.e., about 2200 Pa in about 1000 s. Further shearing does not change the plateau behavior. On the other hand, the 25 mg/ml each P3HT and o-IDTBR solution in o-xylene exhibited a relatively high $G'$ within about 120 s when the temperature was reduced to −5 °C and overall reached its plateau value significantly faster than the P3HT and o-IDTBR blend dissolved in o-DCB solvent. Fig. 14 also shows an phase image of Atomic Force Microscope (AFM) micrograph (captured after the solvent o-DCB is evaporated) of the bulk heterojunction active layer composed of the blend of P3HT with o-IDTBR. The nanostructured nature of the morphology of the blend constituting the active layer is indicated, i.e., the domain sizes are in the 200 nm range.

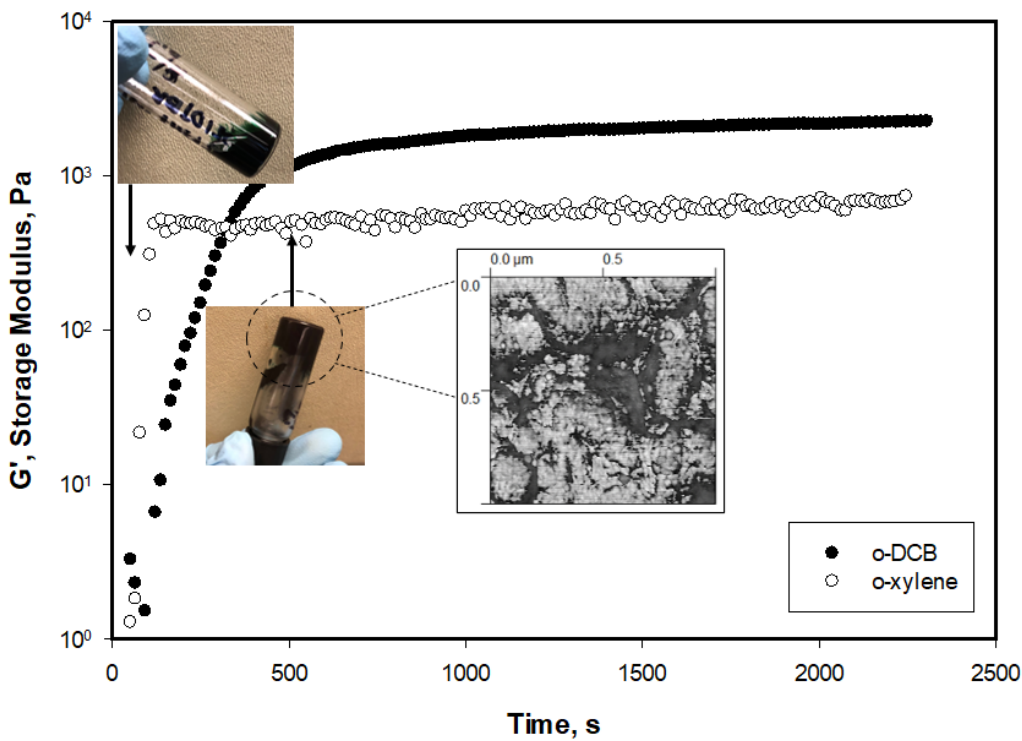

Figure 14. Storage modulus, $G'$, versus time, $t$, of 25 mg/ml P3HT and o-IDTBR solutions in o-DCB and o-xylene at −5 °C.

Fig. 15 displays the comparison of the storage modulus, $G'$, as a function of frequency, $\omega$, after the plateau behavior is reached at −5 °C solutions of pure P3HT and P3HT blended with $PC_{60}BM$ [62. J. He, 2019] and o-IDTBR solutions in two different solvents, i.e., o-DCB and o-xylene. Both the donor and the electron acceptors were kept at a constant concentration of 25 mg/ml. In contrast to the $PC_{60}BM$ making no difference in the gel strength of P3HT for both solvents the replacement of o-IDTBR with $PC_{60}BM$ makes a big difference on the gel strength for both o-DCB and o-xylene. As shown in Fig. 15, the gel strength of the solution is diminished when o-IDTBR is added to P3HT. The reduction of the storage modulus generated by the addition of o-IDTBR is greater for o-DCB (going from $10^4$ Pa for pure P3HT to 2200 Pa for the P3HT/o-IDTBR blend) in comparison to the decrease of the storage modulus with o-xylene (from 1800 Pa for pure P3HT to 700 Pa for the P3HT/o-IDTBR blend).

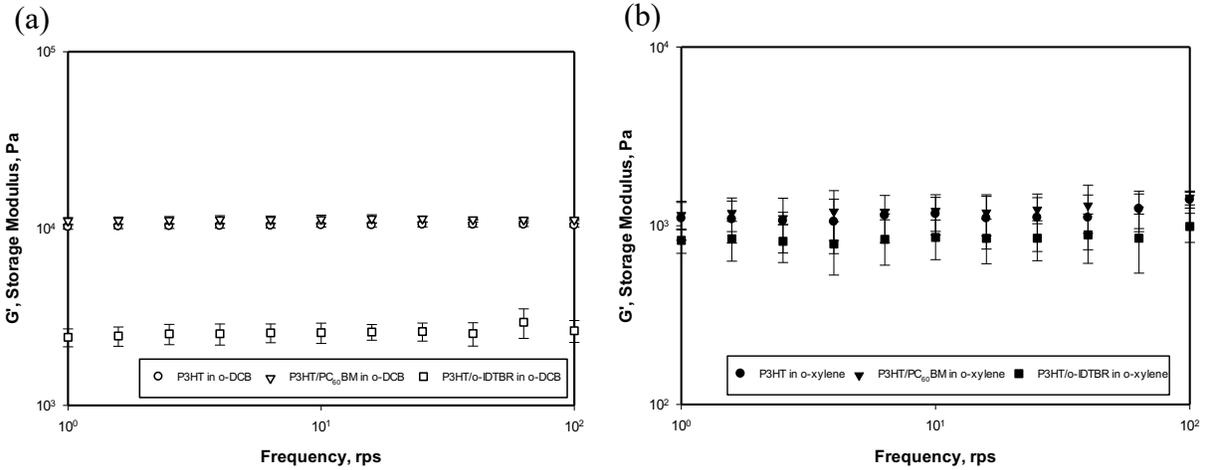

Figure 15. Storage modulus, $G'$, versus frequency, $\omega$, for P3HT, P3HT/PC$_{60}$BM and P3HT/o-IDTBR in two different solvents a) o-DCB and b) o-xylene, all with concentrations of 25 mg/ml.

Fig. 16 displays the storage modulus, $G'$, versus frequency, $\omega$, of P3HT blended with o-IDTBR at different concentration ratios, i.e., 1:1 (25 mg/ml each of P3HT and o-IDTBR), 1:1.6 (25 mg/ml of P3HT and 40 mg/ml o-IDTBR) to 1:2 (25 mg/ml of P3HT and 50 mg/ml o-IDTBR) in o-DCB and o-xylene. The frequency sweeps were again carried out at −5 °C following time sweeps which resulted in reaching of the plateau behavior. Fig. 16 also shows the storage modulus values for pure P3HT at 25 mg/ml. For both concentrations of o-IDTBR blended with P3HT the solution samples exhibited gel behavior upon being cooled to −5 °C and reached a plateau in $G'$. The highest gel strengths were observed for pure P3HT solutions. As indicated earlier, adding o-IDTBR into P3HT system disrupts P3HT crystallization, thus lowering the gel strength. However, these results indicate that the concentration of o-IDTBR in the range investigated did not affect the strength of the gel which forms.

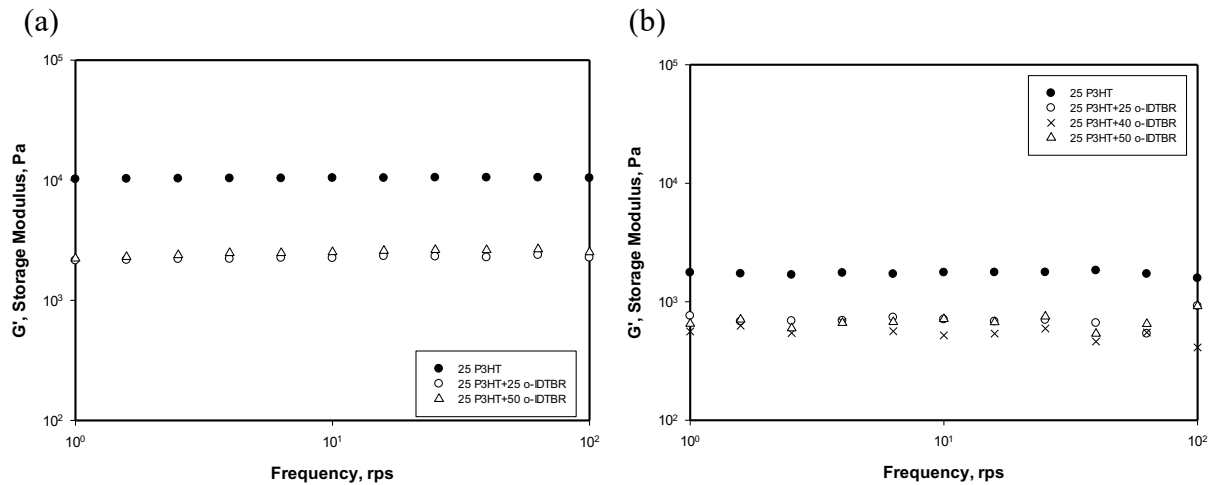

Figure 16. (a) $G'$ versus frequency of solutions comprising blends of P3HT and o-IDTBR in o-DCB with o-IDTBR's concentrations of 0, 25 and 50 mg/ml. (b) $G'$ versus frequency of solutions comprising blends of P3HT and o-IDTBR in o-xylene with o-IDTBR's concentrations of 0, 25, 40 and 50 mg/ml.

The results presented in Fig. 15 and 16 indicate that P3HT and blends of P3HT with $PC_{60}BM$ or with o-IDTBR form gels when subjected to –5 °C. When the solution is held under quiescent conditions or when subjected to small-amplitude oscillatory shearing as a function of time under constant frequency and strain amplitude the structure of the gel evolves until an equilibrium structure is achieved, which coincides with the gel strength remaining constant for longer durations. Previous investigations have shown that the gelation of P3HT when quenched can be considered to occur in two stages. In the first stage the quenching of the solution can result in "viscoelastic phase separation", as affected by the high molecular weight and the associated long relaxation times of the polymer phase, P3HT. The solvent and the electron acceptor whether $PC_{60}BM$ or o-IDTBR exhibit significantly smaller relaxation times. The differences in relaxation behavior following quenching results in the separation of the P3HT from the small molecules to some extent and leads to the formation of solvent holes [59. H. Tanaka, 2000] [60. H. Tanaka, 2006] [61. H. Tanaka, 2009]. The aggregation of the P3HT macromolecules via $\pi$-$\pi$ bonding as explained earlier then leads to the crystallization of the P3HT and the formation of nanowhiskers. The aggregation and the formation of the nanowhiskers generate a network that spans the volume of the sample, giving rise to solid like nature and the gel-like behavior observed via the dynamic properties ($G' >> G''$ and $G' \neq fn(\omega)$).

These results overall show that the nature of the solvent makes a difference in the dynamics of the phase separation and the strength of the gel which forms. Here we have used two solvents with different solubility parameters, i.e., o-DCB with a solubility parameter of 20.1 $MPa^{0.5}$ and o-xylene with a solubility parameter of 18.1 $MPa^{0.5}$ [73. F. Machui, 2012]. The solubility parameter for P3HT is 19.43 $MPa^{0.5}$ [74. S. Lee, 2015]. All the values were measured at 25 °C. The solubility parameter of o-DCB is very close to the solubility parameter of P3HT (20.1 versus 19.43 $MPa^{0.5}$, respectively). On the other hand, the solubility parameter of o-xylene is smaller than that of P3HT (18.1 $MPa^{0.5}$ versus 19.43 $MPa^{0.5}$) indicating that o-DCB is a better solvent for P3HT. The differences are also reflected in the solubility of P3HT in o-DCB and o-xylene. The solubility of P3HT is 14.7 mg/ml in o-DCB, while in o-xylene is 2.7 mg/ml, indicating that o-DCB is indeed a much better solvent whereas o-xylene is a poor solvent (a theta solvent) for P3HT. With a good solvent the P3HT macromolecules exhibit greater radii of gyration (the radius is indicative of the swept volume of P3HT in solvent), whereas with a poor solvent the radius of gyration is much smaller. When the radius of gyration of the macromolecule is reduced in the solution the probability of chains coming into close proximity with each other to give rise to aggregation via $\pi$-$\pi$ bonds and to the crystallization of nanowhiskers to lead to a solid-like network upon the interactions of the nanowhiskers with each other is small, thus reducing the crystallinity and consequently the gel strength. The reduction of the gel strength with the poor solvent o-xylene is clearly shown in Fig. 13-16.

The electron acceptor also plays a significant role, albeit different roles for different acceptors. Incorporation of $PC_{60}BM$ into P3HT does not affect the formation of the gel and the resulting gel strength [61. J. He, 2019] On the other hand, for o-IDTBR the gel strength decreases significantly. These results suggest that o-IDTBR disrupts the aggregation and crystallization of P3HT while $PC_{60}BM$ does not, as shown in Fig. 15 and 16. The mechanism for the disruption should be associated with the intercalation of o-IDTBR into the gallery space between layers of P3HT to disrupt the aggregation and the subsequent crystallization of P3HT [45. Q. Liang, 2019] [75. J. Luke, 2019], suggesting that the P3HT and o-IDTBR are more miscible in each other in comparison with P3HT and PC$_{60}$BM at −5 °C. The mechanisms need to be investigated in subsequent studies. Collins has indicated that PC$_{60}$BM largely migrates into the amorphous region of P3HT molecules. When the concentration of the PC$_{60}$BM is increased beyond a critical value the extra PC$_{60}$BM just phase separates out of the vicinity of P3HT [76. B. A. Collins, 2010].

Fig. 17 shows the results of steady torsional flow for pure P3HT solution, 25 mg/ml of P3HT in o-DCB (Fig. 16a) followed by frequency sweep tests of samples captured at different durations of steady torsional flow (Fig. 16b). The torque, $\mathfrak{I}$, is reported as a function of time when applying a constant apparent shear rate of 1 s$^{-1}$ at −5 °C. The experiment involved quenching the solution specimen to −5 °C and holding the sample under quiescent conditions at −5 °C for 660 s total. Holding the P3HT solution at −5 °C under quiescent conditions gives rise to $G'$ values which are independent of the frequency with $G' >> G''$, i.e., the hallmarks of gel-like behavior. Fig. 17a shows the torque versus time behavior during steady torsional flow following the quiescent holding of the gel at −5 °C for 660 s.

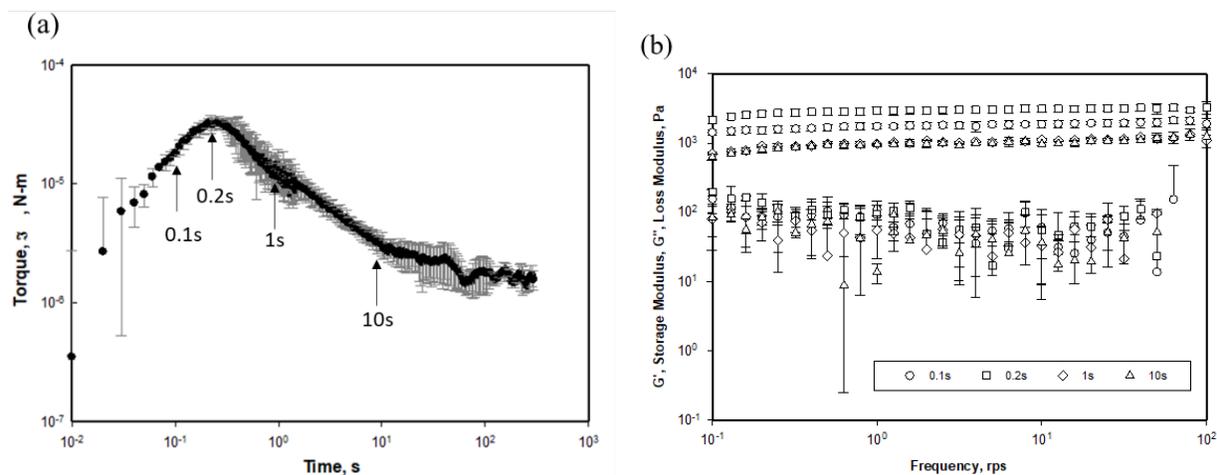

Figure 17. (a) Steady torsional flow of 25 mg/ml P3HT in o-DCB with gap of 0.4 mm at −5 °C.
(b) $G'$ versus frequency, ω, after steady torsional shearing for fixed times,
0.1 s, 0.2 s, 1s and 10s at −5 °C.

The torque versus time result for the P3HT gel during steady torsional flow at −5 °C is very interesting and to our knowledge is being reported for the first time. During steady torsional flow one would have expected the torque (and hence the shear stress) to increase to a constant plateau value and remain there as was observed with the dynamic properties reaching plateau values during oscillatory shear flow. However, this is not what happens. The behavior includes an increase of the torque (hence the shear stress) monotonically until a maximum is reached. The maximum torque is reached at about 0.2 s of shearing. Further shearing during the next 100 s results in the torque to decrease continuously (from a maximum of $3.5 \times 10^{-5}$ N-m at 0.25 s, to $2 \times 10^{-6}$ N-m at 70 s).

Such a decrease can be explained in two ways. The first explanation could be that gels being viscoplastic materials would exhibit wall slip, and the transition from no slip to wall slip can be occurring when a critical shear stress of 250 Pa (torque of $3.5 \times 10^{-5}$ N-m) is reached during shearing at constant apparent shear rate [68. S. Aktas, 2014] [69. J. F Ortega-Avila, 2016] [70. E.

F. Medina-Bañuelos, 2017] [77. E F. Medina-Bañuelos, 2019]. The transition from no slip to wall slip have been observed also with pure polymers, i.e., pure polymer melts can exhibit critical shear stress values at which wall slip is onset. For example, silicone polymers typically start to exhibit wall slip when the shear stress reaches 60 kPa. On the other hand, polyethylene melts typically exhibit a critical shear stress of 200 kPa at which they start to exhibit wall slip [58. D. M. Kalyon, 2004] [78. H. Gevgilili, 2001] [79. D. Kalyon, 2003] [80. E. Birinci, 2006] [81. H. Tang, 2008a] [82. H. Tang, 2008b].

Another mechanism that can explain the observed differences in steady torsional flow versus small-amplitude oscillatory shearing behavior is associated with the evolution of the structure (or lack of it) in the linear versus the non-linear nature of the deformations that are imposed. The time dependent dynamic properties were obtained in the linear viscoelastic region. In the linear region the structure of the gel should not change significantly by definition since both the strain and the strain rate are relatively small, i.e., the structure is not far removed from its equilibrium state. This is predicated on the observation that in the linear region the moduli are independent of the strain amplitude as indeed observed here. On the other hand, the steady torsional flow is being carried out at a relatively high shear rate of 1 s$^{-1}$ which is in the non-linear viscoelastic region so that changes in the structure of the gel may be expected.

A set of supporting oscillatory shearing experiments were carried out to answer the question as to whether it was the wall slip or the structure evolution that gave rise to the torque (and hence the shear stress) versus time behavior noted in Fig. 17. The dynamic properties are very sensitive to the structure of a gel. Thus, inferences on the evolution of the structure of the gel during torsional flow can be made on the basis of the characterization of the dynamic properties following steady torsional flow at constant apparent shear rate of 1 s$^{-1}$ for different durations of time. The methodology involved quenching the sample and holding it at –5 °C for 11 minutes and then the imposition of the steady shear flow and coming to dead stops at various shearing durations, i.e., 0.1, 0.2, 1 and 10 s after the inception of the steady shear flow. The storage and loss moduli as a function of frequency collected on samples that were subjected to shearing in steady torsional flow for different durations are shown in Fig. 8b. First, regardless of the duration of steady torsional flow all samples exhibit gel-like behavior as evidenced by the storage and the loss moduli being independent of the frequency and $G' >> G''$. So the samples remain as gels regardless of the duration of torsional flow. However, the dynamic properties associated with the plateau behavior change significantly with the duration of the steady torsional flow. This indicates that the gel strength depends on the duration of the steady torsional flow that is imposed. The maximum storage modulus, i.e., the maximum gel strength, is observed at 0.2 s, at which a maximum torque is also observed (highest gel strength). The 95% confidence intervals determined according to Student's t-distribution observed for $G'$ are generally narrow indicating the reproducibility of the observed behavior. The $G''$ values exhibit some scatter associated with their relatively small values (about 10-100 Pa) with respect to the transducer sensitivity of the rheometer.

Fig. 17a further shows that following the maximum torque, the storage modulus values decrease significantly (from 3000 Pa at 0.2 s to 1000 Pa at 10 s) as the duration of the steady torsional flow is increased. The gel strengths of solution samples sheared for 1 s are about the same as those sheared for 10 s. This indicates that the structure that was built during steady torsional flow up to the maximum shear stress was subsequently broken in such a way (step change) that additional shearing no longer made a difference on the structure. Thus, at -5 °C there is a critical shear stress for P3HT gels above which the structure of the gel is diminished. This critical shear stress is about 250 Pa (corresponding to the maximum observed torque value of $3.5 \times 10^{-5}$ N-m in Fig. 17a).

Overall, these results indicate that the solutions of P3HT or P3HT blends with electron acceptors become gel-like at $-5$ °C with the gel strength depending on the duration of time that the sample is held under quiescent conditions or under small amplitude oscillatory shearing conditions until a plateau is reached. When this gel is further sheared using steady torsional flow (shearing in the non-linear region) the structure of the gel is altered in such a way that the gel-like behavior prevails but the gel strength is reduced upon shearing beyond the critical shear stress (as indicated by decrease of the storage modulus).

What is the role of wall slip in the observed steady torsional flow behavior? As indicated earlier, for some complex fluids, including polymer melts, wall slip can be onset at a critical shear stress, with the value of the critical shear stress depending on the nature of the polymer melt. This is elucidated next.

Fig. 18 shows the steady torsional flow experiments that were carried out on fresh solutions of 25 mg/ml P3HT in o-DCB at $-5$ °C, following the procedures outlined for the results reported in Fig. 8 but collected using three different gaps, which were 0.4, 0.8 and 1.2 mm. Thus, for each gap the samples were quenched to $-5$ °C, held there for 11 min and then the steady torsional flow was initiated at an apparent shear rate of 1 s$^{-1}$. The overall torque (or shear stress) versus time behavior at $-5$ °C was similar for the three gaps. The torque went up to a maximum, $\mathfrak{J}_{max}$, and then decreased monotonically with continued shearing. The $\mathfrak{J}_{max}$ of these three gaps remained largely unchanged, i.e., in the $3.5 \times 10^{-5}$ N-m to $4.0 \times 10^{-5}$ N-m range. This result indicates that although the value of the maximum torque did not change significantly, the time to reach the maximum torque somewhat increased with the increasing gap values.

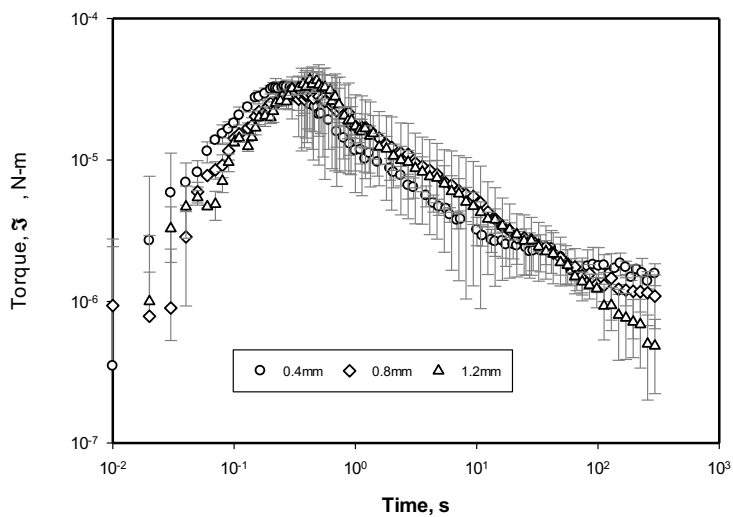

Figure 18. Steady-torsional tests torque, $\mathfrak{J}$, versus time, t, of 25 mg/ml P3HT solutions in o-DCB with gaps varying from 0.4, 0.8, and 1.2 mm at $-5$ °C.

The relatively broad confidence intervals noted precluded any inferences to be made on the effects of the surface to volume ratio of the geometry within which the solution samples are sheared. Generally, with wall slip the shear stress (and hence the torque) decreases with increasing surface to volume ratio (proportional to reciprocal gap). Thus, the true nature of the contribution of wall slip to the observed steady torsional flow behavior could not be made.

DSC measurements were carried out on solutions of blends of P3HT with o-IDTBR first exposed to -5 °C and then drop-cast under ambient temperature conditions in a glove box. During DSC experiments under $N_2$ the temperature was increased from 40 to 300 °C at 5 °C /min (first heating) followed by cooling to 0 °C and then increased to 300 °C again (second heating). The procedure allowed samples to have a consistent thermal history prior to the characterization of their degree of crystallinity. Fig. 19 shows that there are significant differences in the thermal behavior of P3HT versus P3HT blended with o-IDTBR.

During the second heating P3HT/o-IDTBR blend undergoes an exothermic crystallization transition with an onset temperature of 105 °C and a crystallization peak temperature of 117 °C (first peak). With increasing temperature, the crystallites melt starting at a temperature of about around 188 °C and the melting is completed at a temperature of 230 °C. Thus, 230 °C is the highest temperature at which the last trace of crystallinity disappears (definition of the nominal melting temperature). Fig. 19 also shows the thermal behavior of P3HT. P3HT alone does not exhibit the crystallization observed with the P3HT/o-IDTBR blend. Thus, the crystallization peak is associated with the crystallization of the o-IDTBR. With the addition of o-IDTBR, the temperatures for the onset of melting, the peak temperature and the melting temperature (the highest temperature at which the last trace of crystallinity disappears) all shift to lower temperatures in comparison to those encountered for P3HT at the same heating rate, suggesting that the crystallization of P3HT is disrupted by this acceptor. This observation is consistent with the lower gel strengths associated with the P3HT/o-IDTBR blends in comparison to the gel strengths of pure P3HT solutions as shown in Fig. 15 and 16.

For the P3HT samples that are obtained upon thermal treatment at −5 °C followed by solvent evaporation at the ambient temperature, the melting enthalpy (area under the peak) is about 10.1 J/g. Considering that the melting enthalpy of single crystals of P3HT (purely crystalline) is 99.0 J/g [83. J. Zhao, 2009], a degree of crystallinity of about 10.2% is obtained (by weight) indicating that the sample is mostly consisting of the amorphous phase. This emphasizes the importance of the presence of an amorphous phase during the crystallization of P3HT. The migration of the small acceptor molecules into the amorphous regions of P3HT can be substantial [76. B. A. Collins, 2010].

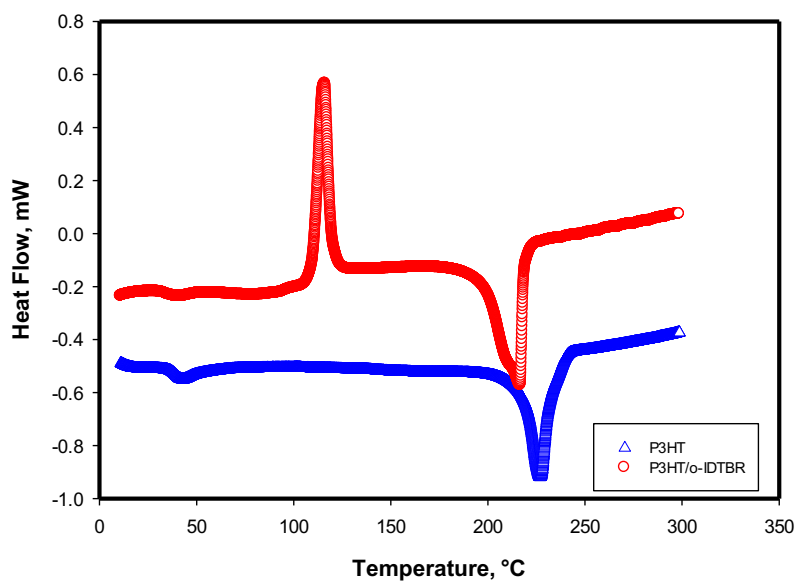

Figure 19. DSC second heating cycles of P3HT and 1:1 o-IDTBR:P3HT blend.

**Conclusions**

A major problem in the fabrication of bulk heterojunction based polymeric solar cells is the lack of a widely available continuous processing method. Generally, spin coating is used to lay a blend of an acceptor with a donor polymer as an active layer film, however, such coating is not scalable. One alternative would be the use of conventional fabrication methods including extrusion and coating of substrates to form the active layer continuously. However, the elasticity and the viscosity of typical acceptor/donor blends are too low to permit extrusion and coating processing under typically ambient conditions. Here we have investigated the gelation of the donor/acceptor blends under sub-ambient conditions at which the blend forms a gel, to exhibit increases in elasticity and viscosity. The gelation dynamics and the associated rheological behavior were investigated for P3HT (regioregular) in conjunction with a non-fullerene acceptor, o-IDTBR and the results were compared with those of P3HT/PCBM blends. Solutions with two different solvents and with various donor/acceptor concentrations were prepared and subjected to sub-ambient conditions (−5 °C). The formation of strong gels of P3HT (as well as blends of P3HT with o-IDTBR) occurred at −5 °C. During the gelation process the dynamic properties increased to reach plateau values. Frequency sweeps were carried out after the plateau behavior was reached and revealed that the dynamic properties became independent of the frequency and that $G' >> G''$. Thus, overall the energy dissipated as heat during oscillatory shearing, i.e., viscous dissipation represented by $G''$ was significantly smaller than the energy stored as elastic energy, as represented by $G'$ (small of tan $\delta = G''/G'$ values) indicating gel like behavior. The time necessary to reach a stable gel structure decreased while the gel strength, i.e., the plateau values of the storage modulus, increased with the increasing concentration of P3HT. Furthermore, the increase of the gel strength of P3HT solutions at −5 °C was particularly significant when the concentration of P3HT increased from 12.5 mg/ml to 25 mg/ml (a jump from 800 to $10^4$ Pa), whereas the gel strength values were relatively low (100-250 Pa) at concentrations which were smaller than 11 mg/ml and less.

The gel strengths of P3HT and P3HT blend solutions increased significantly for a good solvent (o-DCB) and decreased with a poor solvent (o-xylene). This can be linked to the expected increase of the radius of gyration of P3HT in the good solvent, rendering $\pi$-$\pi$ interactions between P3HT macromolecules and hence aggregation more likely, thus increasing the crystallinity of P3HT, giving rise to increased gel strength. The incorporation of the non-fullerene acceptor o-IDTBR into P3HT again led to the formation of strong gels at $-5$ °C but the gel strength was reduced in comparison to that of P3HT blended with PCBM. The concentration of o-IDTBR did not make any difference on gel strength in the 25-50 mg/ml range at a P3HT concentration of 25 mg/ml. In comparison to the conventional $P3HT/PC_{60}BM$ system the gel strengths of P3HT/o-IDTBR blends are smaller, indicating that the crystalline network which forms during   gelation is more pronounced for the P3HT/o-IDTBR system in comparison to $P3HT/PC_{60}BM$.

The time dependence of the steady torsional flows of the gels of P3HT and P3HT blends at $-5$ °C, suggests that shearing decreases the gel strength significantly (independent of the rheometer gap used). This significant dependence of the gel strength on the shearing history is indicative of the narrowness of the processability window of the gels of P3HT and P3HT blends. Shearing reduces the shear viscosity and the elasticity, i.e., the gel strength, of the gels, which form during sub-ambient processing. Overall, these results indicate that the gel structure is fragile in the non-linear viscoelastic range.  This teaches that the rates of shear and elongation imposed during processing of P3HT and P3HT blend solutions in the gel state would play significant roles in the development of gel structures and consequently the functional properties of the active layer.  Thus, the structure of the gel during processing using conventional polymer processing equipment and the subsequent formation of the nanostructure of resulting bulk heterojunction devices, as well as their functional properties including PCEs, would all be interdependent. Such interdependencies will be explored in future investigations.

## Acknowledgments

This work was partially supported by the National Science Foundation under Grant 1635284, funding of the Highly Filled Materials Institute and by a gift from the PSEG Foundation to advance energy innovation at Stevens.  We are grateful for the support received that made this investigation possible.